\documentclass[11pt,twoside]{article}

\usepackage{asp2004}
\usepackage{epsf}
\usepackage{lscape}
\usepackage{graphicx} 

\begin{document}
\title{Search for Giant Planets around White Dwarfs with HST, Spitzer, and VLT}   
\author{S.~Friedrich,$^{1,2}$ H.~Zinnecker,$^{1}$ S. Correia,$^{1}$
  W. Brandner,$^{3}$ M. Burleigh,$^{4}$ and M. McCaughrean$^{5}$ }
\affil{$^1$Astrophysikalisches Institut Potsdam, An der 
    Sternwarte 16, 14482 Potsdam, Germany\\
    $^2$Max-Planck-Institut f\"ur Extraterrestrische Physik, 
    Postfach 1312, 85741 Garching, Germany\\
    $^3$Max-Planck-Institut f\"ur Astronomie, Auf dem
    K\"onigstuhl, 69117 Heidelberg, Germany\\ 
    $^4$Department of Physics \& Astronomy, University of Leicester, 
    Leicester LE1 7RH, UK\\ 
    $^5$School of Physics, University of Exeter, Stocker Road, Exeter EX4 4QL,
    UK\\}

\begin{abstract}
For the last three years we have performed a survey for young ($<3$ Gyrs) giant
planets around nearby white dwarfs with HST, Spitzer, and VLT. 
Direct HST/NICMOS imaging of the seven white dwarfs in the Hyades
gave no evidence for  
companions down to about 10 Jupiter masses and separations larger than
0.5 arcsec ($\sim$\,25\,AU), while VLT/NACO observations revealed
a putative companion to a field white dwarf. 
Second epoch observations with SINFONI on the VLT, however, showed that 
it is most probably a background star. With IRAC on Spitzer we also found no 
indications of cool, very low mass companions in our sample of field white dwarfs.
The implications of these non-detections are briefly discussed.
\end{abstract}


\section{Planets around White Dwarfs?}
Since the discovery of the first extrasolar planet by Mayor \& Queloz
\cite{mayor} the
number of known extrasolar planets has grown to more than 200. Most
discoveries are based on measurements of radial velocities of stellar lines.
Due to the great contrast in optical brightness
between the main sequence star and its planet(s) no planet
has so far been seen directly in the optical range.
A much better spectral contrast between an extrasolar giant planet (EGP)
and a star can be achieved in the IR band, where the planet's thermal
emission peaks. Indeed, recently Chauvin et al. \cite{chauvin}  
published an image of what seems to be the first EGP 
outside our solar system (albeit in a very wide orbit) around the brown
dwarf 2MASSW J1207334-393254 obtained in the infrared with VLT/NACO.
Neuh\"auser et al. \cite{neuhauser} detected a substellar mass object 
around the T Tauri star GQ Lup, also using VLT/NACO, 
and determined a mass between 
1 and 42 Jupiter masses.
The contrast can be increased further if one searches
for planets around {\it white dwarfs}, which are about a factor of ten smaller
than giant planets and typically $10^3$ to $10^4$ times less
luminous than main sequence solar-type stars of spectral type G or K 
(see e.g. Burleigh et al. \citeyear{burleigh}). 

While planets closer than 3--5 AU will probably not 
survive the post-main-sequence evolution and will migrate
inwards and merge (Livio \& Soker \citeyear{livio}), planets 
farther away should survive and adiabatically 
migrate outwards. Their semimajor axis will 
increase by a factor 
$M_{\rm initial}$/$M_{\rm final}$ as the central star loses mass 
(Duncan \& Lissauer \citeyear{duncan}; Burleigh et al. 
\citeyear{burleigh}). The detection of a brown dwarf that survived the 
engulfment by the envelope of a red giant star and migrated inwards
(Maxted et al.\ \citeyear{maxted06}) is a fascinating discovery
in this context and goes to show that brown dwarfs in close orbits
around white dwarfs can exist.

\section{Observations and Data Reduction}
\subsection{HST}
The seven single white dwarfs in the Hyades (HZ4, LB227, VR7, VR16, HZ7, HZ14, LP475-242) 
have been observed with HST/NICMOS in 2003 at 
two roll angles separated by 20 degrees through the F110W and F160W filters. 
Two and four dithered frames, respectively, of 320s integration time each
were recorded for each roll angle and each target in these filters, which
corresponds to a total of 
one and two orbits of HST time per target, respectively. 
The single dithered frames were reduced
using the HST pipeline, then oversampled
by a factor two using cubic spline before registration and a FFT-shift based
combination for each roll angle.
In each filter, the resulting frames corresponding to different roll angles 
were subtracted from each other (see 
Friedrich et al.\ \citeyear{friedrich05} for
details). The results obtained for HZ7 are presented in Fig.~1 as an example.

\begin{figure}[!ht]
\plotfiddle{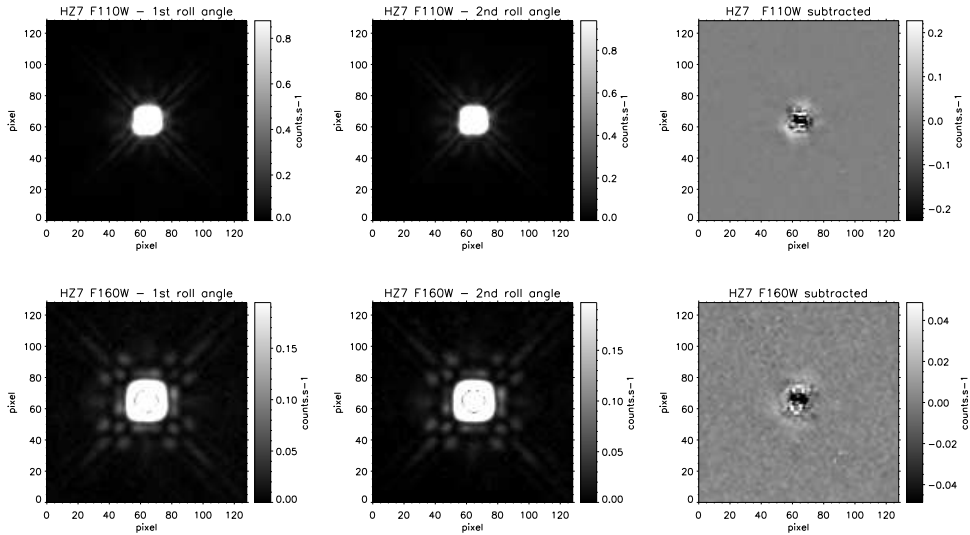}{6.5cm}{90}{50}{50}{190}{-10}
\caption{The HST/NICMOS observations of HZ7 at two different roll angles 
and the result of their subtractions in the F110W filter (upper panel) and
the F160W filter (lower panel). The field of view is about 
$2\farcs75\times 2\farcs75$ (0\farcs0215 pixel size). 
Maximum cuts are at 2\%
of the peak intensity for the observations, and $\pm$0.5\% for the subtracted 
frames}
\end{figure}

\subsection{Spitzer}
Observations were performed with the Spitzer IRAC camera
during a period of two months from
September to November 2005. The nine white dwarfs chosen supplemented
already performed or ongoing programs of similar queries; they
are nearby ($<30$ pc) and younger than 3 Gyrs (cooling plus main 
sequence age). 
With this information mid-IR fluxes of accompanying planets with 
1--10$M_\mathrm{Jup}$ and the expected photometric excess can be predicted 
with models
of Baraffe et al. (\citeyear{baraffe03}). Since these calculations do not include
thermal insolation by or reflected light from the parent star they 
could thus be
regarded as lower limits. Hence it follows that these Spitzer data achieve 
limits of 3--10 M$_\mathrm{JUP}$. 

With standard software packages we processed Basic Calibrated Data (BCD) and 
Post-Basic Calibrated Data (PBCD) and determined fluxes and magnitudes of 
our target stars. According to their effective
temperatures a black body correction was applied (Reach et
al.\ \citeyear{reach05}) and an aperture correction  
for an aperture of 5 pixel radius on source determined (Reach et
al.\ \citeyear{reach05}). Color correction introduces flux changes of about
1--2\%,  aperture correction changes of about 5--7\% depending on the channel. 
Some of the targets 
could not be detected in channel 4. Flux errors for channel 1 and 2 
are on the 1\% level, in channel 3 between 2\% and 4\% and in channel 4
between 5\% and 7\% with two exceptions, where uncertainties in channel 4 
are 14\% and 18\%, respectively. The reason can be found in a higher 
background level.

\subsection{VLT}
In 2004 two field white dwarfs were observed with VLT and NACO in the H band. 
Close to one of our 
targets a second object was found at 0\farcs5 separation 
(see Zinnecker et al.\ \citeyear{zinnecker06}). 
If both objects were physically related, 
this would correspond to a separation of 25 AU at a distance of 
50 pc. The measured absolute $H$ magnitude of 13.7 mag at an 
estimated age of 0.5 Gyr 
yields a mass of $33 M_\mathrm{Jup}$ (Baraffe et al.\ \citeyear{baraffe03}). 
Therefore a second epoch observation was performed in August 2006 with 
VLT and SINFONI in order to obtain information about 
common proper motion and the near-infrared spectrum 
of the putative companion in the $H$- and $K$-band. 
Standard reduction was used 
to produce the image vs.\ wavelength data cube. 

\section{Results}
\subsection{HST}
We validated the method used in the data reduction through simulations. 
For this purpose a fake companion of given brightness was introduced 
in each frame as a scaled version of the target, adding the appropriate noise,
and checking that we could detect the signature of the companion 
after all reduction steps described above.
Fig.~2 shows the results of such a simulation for both filters and the 
same target as in Fig.~1, with a planet at 0\farcs4
separation and 6 magnitudes difference in brightness. One can 
easily notice the pair of negative and positive images of the
companion at the given position (roll angle) from the 
WD-star in the subtracted frame.
Such a simulation 
allows us to estimate the limiting brightness ratio 
as a function of separation and given SNR of the detection. 
According to the COND models of Baraffe et al.\ (2003), 
the limiting mass of the planetary mass companions
that we are able to detect is typically about 10 Jupiter masses 
at 0\farcs5 separation (see Fig.~3 in Friedrich et al.\ 
\citeyear{friedrich05}).
\begin{figure}[!ht]
\plotfiddle{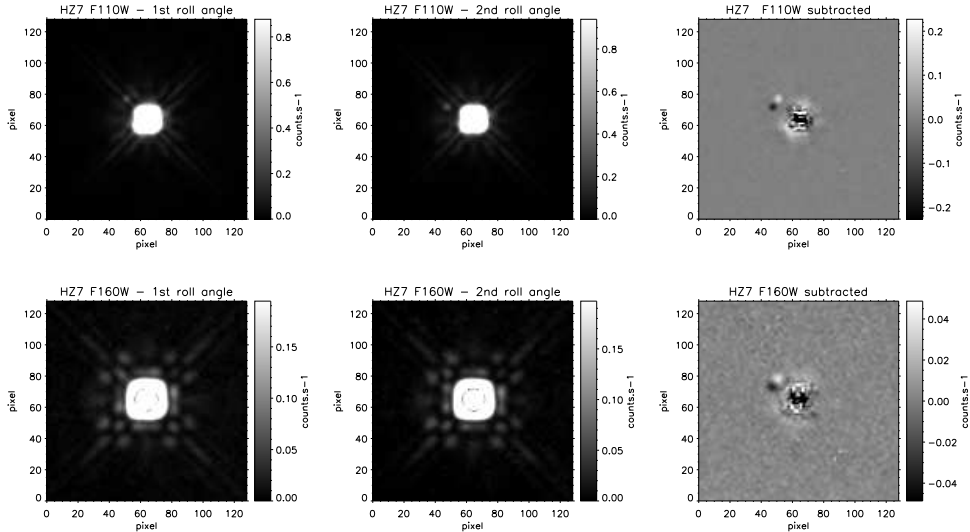}{6.5cm}{90}{50}{50}{190}{-10}
\caption{Same as Fig.~1 but with an artificial planetary mass companion
added (6 mag brightness difference, 0\farcs4 separation)}
\end{figure}

\subsection{Spitzer}
For all white dwarfs, fluxes measured in the four IRAC channels were 
compared to black body fluxes of respective temperatures normalized 
to the $V$ magnitude. 
We are aware that the black body flux will overestimate the white dwarf 
flux in the mid-IR, however, a flux excess in Channel 1 and 2 due to 
a substellar object near the white dwarf would be recognizable in any case.
No excess flux was found for any observed white dwarf in any IRAC channel. 
As an example we show the results for a DA and a DQ white dwarf from 
our sample in Fig.~3.

\begin{figure}[!ht]
\plotone{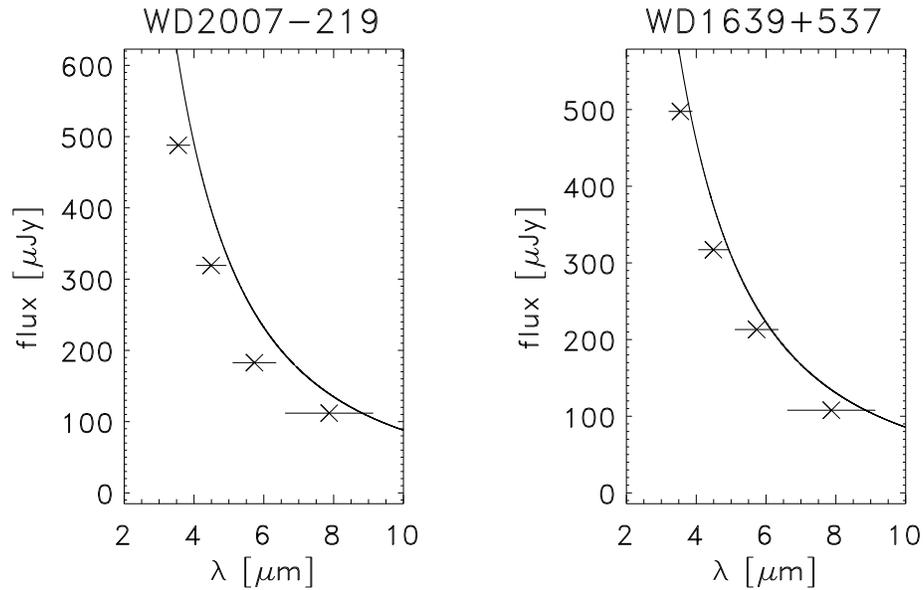}
\caption{Comparison of white dwarf and black body fluxes in 
the four IRAC channels at 3.6$\mu$m, 4.5$\mu$m, 5.8$\mu$m, and 8.0$\mu$m 
for WD2007$-$219 (DA) and WD1639+537 (DQ). The crosses are the fluxes at
the nominal wavelengths of the IRAC bands, horizontal lines 
give the band widths. }
\end{figure}

\subsection{VLT}
In the 2004 and the 2006 observations the position of the white dwarf and 
its putative companion were measured relative to a $\sim$\,0\farcs9 away faint 
($H\sim 20.7$) source detected at both epochs, most probably a field star. 
Whereas the apparent motion of the white dwarf with respect to that object 
is consistent with its known proper-motion, the nearby object does not 
show any significant relative motion relative to the latter (Fig.~4). 
We therefore conclude that it is not physically related to the white dwarf. 
Furthermore, its spectrum is not consistent with it being a very cool 
object. 

\begin{figure}[!ht]
\includegraphics[angle = 270, width=13cm, clip=true]
{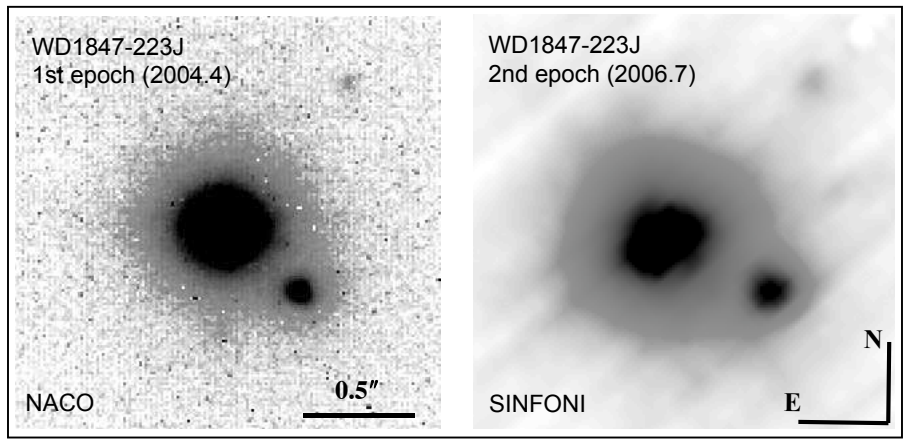} 
\caption{Left panel\,: First-epoch (2004.4) H-band NACO image of the 
field white dwarf WD1847$-$223J with its candidate companion 
at $\sim$\,0$\farcs$5 SW and a faint source at $\sim$\,0$\farcs$9 NW. 
Right panel\,: Second-epoch (2006.7) $H+K$-band SINFONI image 
showing an apparent motion of the white dwarf relative to the two 
sources which is consistent with its known proper-motion 
(110\,mas/yr at PA=117\,$\deg$, east of north, R. D. Scholz, 
private communication).}
\end{figure}

\section{Discussion}
Using available telescopes and focal instruments for infrared 
observations from ground and space, several surveys for planets around 
white dwarfs were performed. Kilic et al.\ \cite{kilic06}, who observed 
18 cool white dwarfs, as well as Farihi et al.\ \cite{farihi05} 
who performed a proper motion and flux excess survey of 261 white dwarfs,  
did not find any evidence for brown dwarfs or planets.
Debes et al.\ (\citeyear{debes06}), who included the result from our 
Hyades observation (Friedrich et al.\ \citeyear{friedrich05}), concluded 
that the upper limit for planets with masses higher than 
$10 M_\mathrm{Jup}$ at separations larger than 23 AU is about 
7\%. So we can ask: Where are the planets? First of all, it must 
be said that the number of white dwarfs in surveys sensitive to planetary 
mass objects, 
although of the order of 100, is still small and that surveys 
are still sensitivity limited such that $1 M_\mathrm{Jup}$ planets 
cannot be detected. Nevertheless it appears that either 
massive Jupiter-like planets in wide orbits around 3--4 M$_\odot$
main sequence progenitor stars are not common, that
planets do not survive the post-main sequence evolution or that 
massive Jupiter-like planets do not remain bound due to planet-planet 
perturbations (Debes \& Sigurdsson \citeyear{debes}).

\acknowledgements 
We would like to thank R. Napiwotzki for making his white dwarf data available 
to us and R.D. Scholz for perfoming 
an accurate determination of the proper-motion of the field white dwarf 
WD1847$-$223J. 


\end{document}